# Evaluating CRM Implementation in Healthcare Organization

Muhammad Anshari [1], and Mohammad Nabil Almunawar [2+]

[1, 2] University of Brunei Darussalam

**Abstract.** Recently, many healthcare organizations are adopting CRM as a strategy, which involves using technology to organize, automate, and coordinate business processes, in managing interactions with their patients. CRM with the Web technology provides healthcare providers the ability to broaden their services beyond usual practices, and thus offers suitable environment using latest technology to achieve superb patient care. This paper discusses and demonstrates how a new approach in CRM based on Web 2.0 will help the healthcare providers improving their customer support, avoiding conflict, and promoting better health to patient. With this new approach patients will benefit from the customized personal service with full information access to perform self managed their own health. It also helps healthcare providers retaining the right customer. A conceptual framework of the new approach will be discussed.

## 1. Introduction

One of the most interesting aspects in medical care is how to manage the relationship between healthcare providers and patients. Fostering relationship leads to maintain loyal customer, greater mutual understanding, trust, patient satisfaction, and patient involvement in decision making (Richard and Ronald, 2008). Furthermore, effective communication is often associated with improved physical health, more effective chronic disease management, and better health related quality of life (Arora, 2003). On the other hand, failure in managing the relationship will affect to the patient dissatisfaction, distrust towards systems, patient feels alienated in the hospital, and jeopardize business survivability in the future.

This paper is motivated by the following case between a patient whose name is Prita Mulyasari (Prita) with Omni International Hospital Tangerang, Indonesia (Omni). The case generated massive public attention channeled through various media, including social network sites. We will use this case to motivate us in developing a new CRM model to address relationship management between patients and healthcare providers that incorporate the latest development in Information and Communication Technology (ICT).

*Prita Mulyasari is a housewife and mother of two who was a patient at Omni International Hospital, Tangerang Jakarta for an illness that was eventually misdiagnosed as mumps. Her complaints and dissatisfaction about her treatment which started as a private email to her friends in September 2008 were made public rapidly distributed across forums via online mailing lists. Once the email became public knowledge, Omni responded by filing a criminal complaint and a civil lawsuit against Prita. Then, verdict against Prita, at Banten District Court on May 13, 2009, she was sentenced to six years jail and fined 204 million rupiah (US$ 20,500). Support from a group on Facebook has attracted considered support as well as Indonesian Blog. A mailing list and Facebook group called "KOIN UNTUK PRITA" (Coins for Prita) started raising money from people throughout Indonesia. People began collecting coins to help Prita to pay the fine. Seeing the huge support for Prita, Omni International Hospital dropped the civil lawsuit. Significant pressure eventually led to Prita being released from detention on June 3, 2009 (Detik.com, 2008; thejakartapost.com, 2009).*

---
+ Dr.Hj.Mohd.Nabil Almunawar, FBEPS-UBD, Jl.Tungku Link Gadong, BE 1410 Brunei Darussalam. Tel.: + 673 8825711; fax: +673 2465017. *E-mail address*: nabil.almunawar@ubd.edu.bn / Muhammad Anshari, State Islamic University Yogyakarta Indonesia *E-mail address*: anshari@yahoo.com.

The main goal of this paper is to introduce a promising future research direction which will shape the future of health informatics. In this paper we will discuss and demonstrate how the new approach will help the healthcare increasing their customer support, avoiding conflict, and promoting better health to patient. A conceptual framework of such new approach will be proposed. The structure of the paper is as follows. In sections 2, we discuss recent CRM development. Customer Service and Value Creation in healthcare are discussed in Section 3 and 4 respectively, and finally we propose our conceptual model for CRM in healthcare in Section 5.

## 2. Customer Relationship Management (CRM)

Currently, a new paradigm has appeared in CRM systems as a result of the development of IT and Web service. This new paradigm has been named as Social CRM or CRM 2.0 (Greenberg, 2009) because it is based on Web 2.0. That is more focused on the conversation that is going on between healthcare-patient and patients-others. The term of Social CRM and CRM 2.0 is used interchangeably. Both share new special capabilities of social media and social networks that provide powerful new approaches to surpass traditional CRM. Denis Pombriant from Beagle Research Group (2009) analyzed that customers has more control of their relationships with organization a lot more than they were in few years ago. It is because customers have accessed to new levels of education, wealth and information. The change is a social change that impacts all organizations.

The Prita's case mentioned in the introduction is a good example how CRM 2.0 will be useful in the context of Healthcare's service. Prita shared her experiences about the bad service received from the hospital to her friends through a mailing list. Unfortunately, the hospital sued Prita and took the case to the court for the defamation. The responds from other users (public) were immense, they viewed that the hospital was arrogant and the case should be withdrawn. Facebook users collaborated to help Prita by collecting coin paying her fines to the hospital. View the case from CRM 2.0 perspective, the hospital failed to understand that the type of relationship, connection and generated value have changed in dealing with the customers. As a matter of fact, Prita established communication not only with the hospital by complaining about her dissatisfaction but also with her friends on the social network. In this scenario, the hospital cannot control the conversation between patients with others because ICT enables them to share with others without barriers. Public has built image from the effect of conversation taken place on social network that the hospital unprofessional in conducting medical activities which led to distrust to the service.

## 3. Customer Service in Healthcare

As a business, healthcare organization stands in need of the same standards of customer service as other industries or business organizations. The fact that customer service expectations in healthcare organization are high poses a serious challenge for healthcare providers as they have to make exceptional impression on every customer. Healthcare organizations strategies should transform customer strategies and systems to customer engagement. Proactive strategies will improve customer services. And great customer support will increase loyalty, revenue, brand recognition, and business opportunity.

Reacting to problems after they happen is usually more expensive than addressing them proactively. The case of Prita provides a good lesson that the healthcare organization was not able to manage the relationship with their patients, Moreover, the hospital should be able to manage conversation with the patients which is missing in the case of CRM implemented by the hospital.

## 4. Value Creation in Healthcare

It is important to consider that each business process as a layer of value to the service. Patients place a value on these services according to quality of outcome, quality of service, and price. The value of each layer depends on how well they are performed. When a healthcare provider cannot achieve its strategic objectives, it needs to reengineer their activities to fit business processes with strategy. If the business processes unfit with strategy, it will diminish the value. For example, the value of a health education is reduced by a delay respond of patient's query or poor communication skills. The value of service is reduced by a poor schedule of physician. Porter (1985) proposed value chain framework for the analysis of organizational level competitive strengths and weaknesses. Further research conducted by Stabell and Fjeldstad (1998) refined the three distinct generic value configuration models (value chain, value shop, and value network) required to understand and analyze firm-level value creation logic across broad range of industries. The value shop diagram will be used to evaluate Prita's case shaping proposed framework.

## 5. The Proposed Social CRM Model

The model operates in the area of healthcare organization–patient relationships inclusive with social networks interaction, and how they possibly shared information to achieve health outcomes. Figure 1 is a proposed model of Social CRM in healthcare environment. It offers a starting point for identifying possible theoretical mechanisms that might account for ways in which Social CRM provides one-stop service for building relationship between healthcare organization, patients, and community at large.

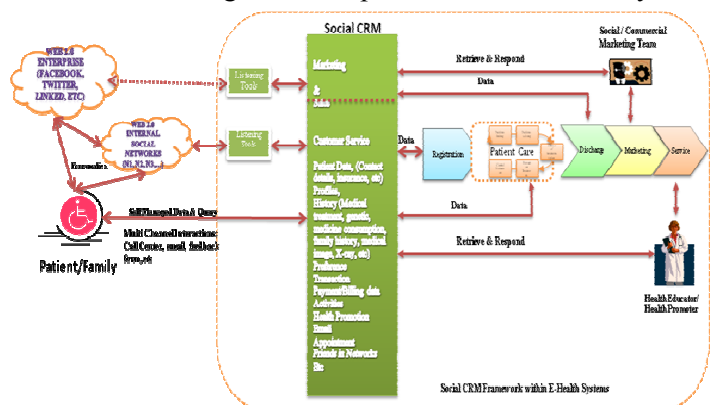

Fig. 1: Proposed Model of Social CRM in Healthcare.

The framework is developed from Enterprise Social Networks, Internal Social Networks, Listening tool interfaces, Social CRM systems within healthcare provider, and healthcare value configuration (value chain and value shop).

Social Networks refers to the Web 2.0 technology that patient or his/her families may join any of them. It differentiates two social networks linkages to the patient or his/her family; they are Enterprises Social Networks and Internal Social Networks. The Enterprises Social Networks refers to external and popular Web 2.0 applications such as Facebook, Twitter, LinkedIn, MySpace, Friendster, etc which patient may belong to any of those for interaction. The dashed line connected enterprises social networks and CRM systems mean that none of those networks have control over the others directly, but constructive conversation and information from enterprises social networks should be captured for creating strategy, innovation, better service and at the same time responds accurately.

Lesson learnt from Prita's case, the hospital was not proficient in capturing the message from customers at social networks because they did not consider it in their relationship management strategy that the customers have changed and they made conversation, judging hospital's value, criticizing their services at those networks which led to distrust towards the hospital's allegation and jeopardized the business in the long run.

Further, it proposes Internal Social Networks that operated, managed, and maintained within healthcare's infrastructure. This is more targeted to internal patients/families within the healthcare to have conversation between patients/family within the same interest or health problem/ illness. For example, patient with diabetic would motivate to share his/her experiences, learning, and knowledge with other diabetic patients. Since patient/family who generates the contents of the Web, it can promote useful learning center for others, not only promoting health among each others, but also it could be the best place supporting group and sharing their experiences related to all issues such as; how the healthcare does a treatment, how much it will cost them, what insurance accepted by healthcare, how is the food and nutrition provided, etc. Therefore, this is generic group that will grow depends on the need of patients in that healthcare. For instant, N1 is internal social networks for Diabetic, N2 is for Cancer, N3 is for hearth disease, and so on. Creating Internal Social

Networks is part of the strategy to isolating problem into small space or more focus to the local's problem so it can be easily monitored and solved. Moreover, this strategy will promote loyalty of customers to keep using service from the healthcare.

In general, the aim to put together linkage of internal and external social networks are to engage patients and export ideas, foster innovations of new services, quick response/feedback for existing service, and technologies from people inside and outside organization. Both provide a range of roles for patient or his/her family. The relationships can create emotional support, substantial aid and service, influence, advice, and information that a person can use to deal with a problem. In addition, listening tool between Social Networks and CRM systems is mechanism to capture actual data from social media and propagates this information forward to the CRM. This tool should be capable to filter noise (level of necessity for business process) from actual data that needs to be communicated to CRM.

Since this strategy is absent in the Prita's case, the hospital was late to realize that the patient dissatisfied with the service from beginning, and the hospital assumed that everything was fine, until she communicated her dissatisfaction through her social networks. Responding this problem, the hospital should isolate the internal problem like dissatisfaction of patient by quick response to resolve the issue before it gets bigger and uncontrolled. The Internal Social Networks could be solution to prevent the same problem in the future.

Social CRM empowers patient/family to have the ability in controlling his own data. Once patient/family registers to have service from healthcare provider, it will enable them to have personalized e-health systems with Social CRM as the frontline of the system. The system will authorize for each patient then; the authorization and self-managed account/service are granted to access all applications and data offered by the systems. This authorization is expected to be in the long run since the information and contents continue to grow. Technical assistant is available through manual or health informatics officer (just like any other customer service in business/organization) who stand by online assisting patient/family in utilizing the systems especially for the first timer. Furthermore, since all the information (medical records) can be accessed online everywhere and anytime, it will enable collaborative treatment from telemedicine.

Consider this scenario; while we go to physician for diagnose, sometimes there is trade-off between time allocated each patient and comprehensiveness of diagnosing process. Long queue patient waits for consultation make healthcare provider to be able to allocate time wisely for each patient. Within the constraint of consultation time yet physician is able to conduct diagnosis efficiently and effectively. The system supports the customer service because it helps both healthcare provider and patient in diagnose activity. The physician will have complete information, knowledge, and saving a lot of time to learn about patient history because patient participate in the detailing his medical records data through the system, and patient benefits from quality of diagnoses' time because his medical records are overviewed in full scene. In other words, it can provide better customer service to meet patient's expectation and improve quality of consultation time. The physician is expected to have comprehensive view of the patient's history before diagnosing or analyzing consulted symptoms. This can be achieved because physician will be able to observe the report of patient's medical history such as last medicine consumption, previous diagnoses, lab result, activities suggested by health educator, etc. In addition, by empowering patients with medical data and personalized e-health, the healthcare needs to provide officer in duty (health educator/ health promoter) in order to interpret medical data or respond online query/consultation. The officer in duty is required to have an ability to interpret medical data and also familiar with the technical details of the systems.

Social CRM functionalities are composed from Marketing, Sales, and Customer Service which operated to achieve business strategy of healthcare organization. For example, marketing's strategy should accommodate social marketing to promote public health and commercial marketing to acquire more customers coming for services. Customer service will offer distinct value for each activity. The different from the traditional CRM, the state to empowering for self managed data and authorization will encourage patient willingly to provide full data without hesitation. More data provided more information available for the sake of analyzing for the interest of marketing, sales, and customer service.

The own unique characteristics value creation by adopting Social CRM is able to generate contents from both parties either from healthcare and also patients. Social CRM in healthcare is providing value-added services to patients like openness of medical records, improving patient loyalty, creating better healthcare-patient communication, improving brand image and recognition, and self managed data which will improve health literacy to reduce economic burden for society to the whole.

The framework above proposes and combines the concept of value chain and value shop. As discussed in previous section, the raw data arrives in one state, and leave in another state. The patient enters ill and leaves well (hopefully). The activities of value chain are; arriving from registration, patient care, discharge, marketing, and service—producing data at respective state. However, in the process of patient care is elaborated according to the value shop where value is created by mobilizing resources and activities to resolve a particular patient's problem. The five generic categories of primary value shop activities; Problem-finding and acquisition, choosing the overall approach to solving the problem execution, and control & evaluation.

Furthermore, the framework accommodates Social/Commercial Marketing team. Social marketing is more prevalence to the government healthcare that operates as an agent of the public at large. Campaign of healthy life for the healthcare is example of Social Marketing. It is not intended for commercial benefit for short term but it is beneficial for the community. On the other hand, commercial marketing is standard marketing strategy exist for any business entities. Both are acting in responds to the public demands like social networks, Mailing list, blog, etc.

The other feature of the model is robustness of systems because more applications/services will be added as characteristics of Web 2.0. Some of the features that available to the user are; updating personal data, Medical Records & History (medical treatment received, medicine consumption history, family illness history, genetic, medical imaging, x-ray, etc), Preference services, Transaction, Payment/Billing data, Activities, Personal Health Promotion and Education, Email, Appointment, Friend in networks, forums, chatting, etc.

The adoption of Social CRM to healthcare provider prevents any dispute and avoiding conflict between healthcare organization and patient. Prita's case in introduction took place because; the hospital needs to understand that behavior and expectation of patients continue to change eventually. We propose that Social CRM framework as alternative solution to the hospital. Once the case became a public knowledge, it affects survival ability of the hospital in the long run jeopardize due to loosing of the trust towards the service. Therefore, the hospital should perform re-engineering process to adapt their CRM strategy and tool in order to acquire potential customer coming for the service. The framework will give prospect to acquire, retain, and extend relationship with their patients.